\documentstyle[12pt]{article}

\begin{document}

\begin{flushright}
GUTPA/01/04/02
\end{flushright}
\vskip .1in

\begin{center}
{\Large {\bf Lorentz Invariance and Origin of Symmetries }}

\vspace{35pt}

{\bf J.L. Chkareuli}

\vspace{6pt}

{\em Institute of Physics, Georgian Academy of Sciences, 380077 Tbilisi,
Georgia\\[0pt]
}

\vspace{12pt}

{\bf C.D. Froggatt}

\vspace{6pt}

{\em Department of Physics and Astronomy\\[0pt]
Glasgow University, Glasgow G12 8QQ, Scotland\\[0pt]
}

\vspace{12pt}

{\bf H.B. Nielsen} 
\vspace{6pt}

{\em Niels Bohr Institute, \\[0pt]
Blegdamsvej 17-21, DK 2100 Copenhagen, Denmark}

\bigskip

{\large {\bf Abstract}}
\end{center}

In this letter we reconsider the role of Lorentz invariance in the dynamical
generation of the observed internal symmetries. We argue that, generally,
Lorentz invariance can only be imposed in the sense that all Lorentz
non-invariant effects caused by the spontaneous breakdown of Lorentz
symmetry are physically unobservable. Remarkably, the application of this
principle to the most general relativistically invariant Lagrangian, with
arbitrary couplings for all the fields involved, leads by itself to the
appearance of a symmetry and, what is more, to the massless vector fields
gauging this symmetry in both Abelian and non-Abelian cases. 
In contrast, purely global symmetries are only generated as 
accidental consequences of the gauge symmetry.

\thispagestyle{empty} \newpage

It is still a very attractive idea that a local symmetry for all the
fundamental interactions of matter and the corresponding massless gauge
fields could be dynamically generated (see \cite{book} and extended
references therein). In particular there has been considerable interest 
\cite{bj} in the interpretation of gauge fields 
as composite Nambu-Jona-Lasinio
(NJL) bosons \cite{njl}, possibly associated with the spontaneous breakdown
of Lorentz symmetry (SBLS). However, in contrast to the belief advocated in
the pioneering works \cite{bj}, there is a generic problem in turning
the composite vector particles into genuine massless gauge bosons \cite
{suzuki}.

In this note we would like to return to the role of Lorentz symmetry in a
dynamical generation of gauge invariance. We argue that, generally, Lorentz
invariance can only be imposed in the sense that all Lorentz non-invariant
effects caused by its spontaneous breakdown are physically unobservable. We
show here that the physical non-observability of the SBLS, taken as a basic
principle, leads to genuine gauge invariant theories, both Abelian and
non-Abelian, even though one starts from an arbitrary relativistically
invariant Lagrangian. In the original Lagrangian, the vector fields are
taken as massive and all possible kinetic and interaction terms are
included. However, when SBLS occurs and its non-observability is imposed,
the vector bosons become massless and the only surviving interaction terms
are those allowed by the corresponding gauge symmetry. Thus, the Lorentz
symmetry breaking does not manifest itself in any physical way, due to the
generated gauge symmetry converting the SBLS into gauge degrees of freedom
of the massless vector bosons. Remarkably, even global symmetries are not
required in the original Lagrangian---the SBLS induces them automatically as
accidental symmetries accompanying the generated gauge theory.

In order to consider general interactions between a vector field and
fermionic matter, it is convenient to use 2-component left-handed Weyl
fields $\psi _{Li}$ to represent the fermions. For simplicity we shall
consider the case of two Weyl fields ($i=1,2$), which will finally be
combined to form a Dirac-like field $\psi $ = 
${{\psi _{L1} \choose \psi _{L2}^{\dagger }}}$ in Weyl representation.

The most general Lagrangian density, 
only having terms of mass dimension 4 or less, for
a theory containing a pure spin-1 vector field and two Weyl fermions is:
\begin{eqnarray}
L(A,\psi ) &=&-\frac{1}{4}F_{\mu \nu }F_{\mu \nu }+\frac{1}{2}M^{2}A_{\mu
}^{2}+i\sum_{j=1}^{2}\psi _{Lj}^{\dagger }\sigma _{\mu }\partial _{\mu }\psi
_{Lj}  \nonumber \\
&&-\sum_{j,k=1}^{2}\epsilon ^{\alpha \beta }
(m_{jk}\psi _{Lj\alpha }\psi _{Lk\beta } + m_{jk}^{\star }
\psi _{Lj\alpha }^{\dagger }\psi_{Lk\beta }^{\dagger })  \nonumber \\
&&+\sum_{j,k=1}^{2}e_{jk}A_{\mu }\psi _{Lj}^{\dagger }\sigma _{\mu }\psi
_{Lk}+\frac{f}{4}A_{\mu }^{2}\cdot A_{\mu }^{2}  \label{L}
\end{eqnarray}
with the Lorentz condition ($\partial _{\mu }A_{\mu }=0$) imposed as an
off-shell constraint, singling out a genuine spin-1 component in the
four-vector $A_{\mu }$. 
This constraint also ensures that, after an appropriate scaling, 
the kinetic term for $A_{\mu}$ can be written in the 
usual $-\frac{1}{4}F_{\mu \nu }F_{\mu \nu}$ form.  
Note that $m_{jk}=m_{kj}$, as a consequence of 
Fermi statistics. It is always possible to simplify this 
Lagrangian density, by defining two new left-handed Weyl spinor fields which
transform the ``charge term'' $\sum_{j,k=1}^{2}e_{jk}A_{\mu }\psi
_{Lj}^{\dagger }\sigma _{\mu }\psi _{Lk}$ into the diagonal form $%
\sum_{k=1}^{2}e_{k}A_{\mu }\psi _{Lk}^{\dagger }\sigma _{\mu }\psi _{Lk}$.

Let us consider now the SBLS in some detail. We propose that the vector
field $A_{\mu }$ takes the form
\begin{equation}
A_{\mu }=a_{\mu }(x)+n_{\mu }  \label{f}
\end{equation}
when the SBLS occurs. Here the constant Lorentz four-vector $n_{\mu }$ is a
classical background field appearing when the vector field $A_{\mu }$
develops a vacuum expectation value (VEV).
Substitution of the form (\ref{f}) into the Lagrangian (\ref{L}) immediately
shows that the kinetic term for the vector field $A_{\mu }$ translates into
a kinetic term for $a_{\mu }$ ($F_{\mu \nu }^{(A)}=F_{\mu \nu }^{(a)}$),
while its mass and interaction terms are correspondingly changed. As to the
interaction term, one can always make a unitary transformation to two new
Weyl fermion fields ${\bf \Psi }_{Lk}$
\begin{equation}
\psi _{Lk}=\exp [ie_{k}\omega (x)]\textrm{ }{\bf \Psi }_{Lk}\ 
\textrm{, \qquad}
\omega (x)=n\cdot x  \label{psi}
\end{equation}
so that the Lorentz symmetry-breaking term $n_{\mu }\cdot
\sum_{k=1}^{2}e_{k}\psi _{Lk}^{\dagger }\sigma _{\mu }\psi _{Lk}$ is exactly
cancelled in the Lagrangian density $L(a_{\mu }+n_{\mu },\psi ).$ This
cancellation occurs due to the appearance of a compensating term from the
fermion kinetic term, provided that the phase function $\omega (x)$ is
chosen to be linear\cite{ferrari} in the coordinate four-vector $x_{\mu }$
(as indicated in Eq.~\ref{psi}). However, in general, the mass terms will
also be changed under the transformation (\ref{psi}):
\begin{equation}
m_{jk}\psi _{Lj\alpha }\psi _{Lk\beta }\rightarrow m_{jk}\exp \left[
i(e_{j}+e_{k})n\cdot x\right] {\bf \Psi }_{Lj\alpha }{\bf \Psi }_{Lk\beta }
\label{mass}
\end{equation}
If $e_{j}+e_{k}\neq 0$ for some non-zero mass matrix element $m_{jk}$, the
transformed mass term will manifestly depend on $n_{\mu }$ through the
translational non-invariant factor $\exp \left[ i(e_{j}+e_{k})n\cdot
x\right] $, which in turn will visibly violate Lorentz symmetry. So our main
assumption of the unobservability of SBLS implies that we can only have a
non-zero value for $m_{jk}$ when $e_{j}+e_{k}=0$.

After imposing these conditions on the charges, the remaining traces of SBLS
are contained in the vector field mass term and the $A_{\mu }^{2}\cdot
A_{\mu }^{2}$ term. Thus the remaining condition for the non-observability
of SBLS becomes:
\begin{equation}
\lbrack M^{2}\textrm{ }+f(a^{2}+(n\cdot a)+n^{2})](n\cdot a)=0  \label{cons**}
\end{equation}
An extra gauge condition $n\cdot a$ $\equiv n_{\mu }\cdot a_{\mu }=0$ would
be incompatible with the Lorentz gauge ($\partial _{\mu }A_{\mu }=0$) 
already imposed on the vector field $a_{\mu }$. Therefore, the only
way to satisfy Eq.~(\ref{cons**}) is to take $M^{2}=0$ and $f=0$. Otherwise
it would either represent an extra gauge condition on $a_{\mu }$, or it
would impose another dynamical equation in addition to the usual Euler
equation for $a_{\mu }$.

Thus imposing the non-observability of SBLS, the Lorentz gauge restriction
and the presence of terms of only dimension 4 or less
has led us to the Lagrangian density for chiral
electrodynamics, having the form:
\begin{eqnarray}
L &=&-\frac{1}{4}F_{\mu \nu }F_{\mu \nu }+i\sum_{k=1}^{2}{\bf \Psi ^{\dagger
}}_{Lk}\sigma _{\mu }(\partial _{\mu }-ie_{k}a_{\mu }){\bf \Psi }_{Lk}
\nonumber \\
&&-\sum_{j,k=1}^{2}(m_{jk}{\bf \Psi }_{Lj\alpha }{\bf \Psi }_{Lk\beta
}\epsilon ^{\alpha \beta }+h.c.)  \label{Lcqed}
\end{eqnarray}
with the restriction that $m_{jk}=0$ unless $e_{j}+e_{k}=0$. In general,
i.e.~when $\sum_{k}e_{k}^{3}\neq 0$, even this Lagrangian density will lead
to the observability of the SBLS, because of the presence of
Adler-Bell-Jackiw anomalies \cite{ABJ} in the conservation equation for the
current $j_{\mu }^{A}=\sum_{k}e_{k}{\bf \Psi ^{\dagger }}_{Lk}\sigma _{\mu}
{\bf \Psi }_{Lk}$ coupled to $A_{\mu }$. We are now interpreting the ${\bf 
\Psi }_{Lk}$ as the physical fermion fields. However, in momentum
representation, the transformation (\ref{psi}) corresponds to displacing the
momentum of each fermion by an amount $e_{k}n_{\mu }$ This induces a
breakdown of momentum conservation, which can only be kept unobservable as
long as the charge associated with the current $j_{\mu }^{A}$ is conserved.
This means that an anomaly in the current conservation will also violate
momentum conservation by terms proportional to $n_{\mu }$. Such a breaking
of momentum conservation would also give observable Lorentz symmetry
violation. So the only way to satisfy our non-observabilty of SBLS principle
is to require that the no gauge anomaly condition
\begin{equation}
\sum_{k}e_{k}^{3}=0  \label{anomaly}
\end{equation}
be fulfilled. For the simple case of just two Weyl fields, this means that
the two charges must be of equal magnitude and opposite sign, 
$e_{1}+e_{2}=0$. This is also precisely the condition that must be satisfied
for a non-zero mass matrix element $m_{12}=m_{21}\neq 0$. If the charges are
non-zero, the diagonal (Majorana) mass matrix elements vanish, 
$m_{11}=m_{22}=0$, and the two Weyl fields correspond to a massive 
particle described by the Dirac field ${\bf \Psi }$ = 
${{\bf \Psi }_{L1} \choose {\bf \Psi }_{L2}^{\dagger }}$. 
Thus we finally arrive at gauge invariant QED as the only version of the
theory which is compatible with physical Lorentz invariance when SBLS 
occurs.

Let us now consider the many-vector field case which can result in a
non-Abelian gauge symmetry. We suppose there are a set of pure spin-1
vector fields, $A_{\mu }^{i}(x)$ with $i=1,...N$, 
satisfying the Lorentz gauge condition, but not even proposing 
a global symmetry at the start. The matter fields 
are collected in another set of Dirac fields $\psi
=(\psi ^{(1)},...,\psi ^{(r)})$. Here, for simplicity, we shall neglect
terms violating fermion number and parity conservation. The general
Lagrangian density $L(A_{\mu }^{i},\psi)$ describing all 
their interactions is given by:
\begin{eqnarray}
L = -\frac{1}{4}F_{\mu \nu }^{i}F_{\mu \nu}^{i}
+\frac{1}{2}(M^{2})_{ij}A_{\mu }^{i}A_{\mu }^{j}
+\alpha ^{ijk}\partial _{\nu }A_{\mu
}^{i}\cdot A_{\mu }^{j}A_{\nu }^{k} &&  \nonumber \\
+\beta ^{ijkl}A_{\mu }^{i}A_{\nu }^{j}A_{\mu }^{k}A_{\nu }^{l}+i
\overline{\psi }\gamma \partial \psi -\overline{\psi }m\psi +A_{\mu }^{i}
\overline{\psi }\gamma _{\mu }T^{i}\psi &&  \label{LN}
\end{eqnarray}
Here $F_{\mu \nu }^{i}=\partial _{\mu }A_{\nu }^{i}-\partial _{\nu }A_{\mu
}^{i}$ , 
while $(M^{2})_{ij}$ is a general $N\times N$ mass-matrix for the vector
fields and $\alpha ^{ijk}$ and $\beta ^{ijkl}$ are dimensionless coupling
constants. The $r\times r$ matrices $m$ and $T^{i}$ 
contain the still arbitrary fermion masses and
coupling constants describing the interaction between the fermions and the
vector fields (all the numbers mentioned are real and the matrices
Hermitian, as follows in this case from the Hermiticity of the Lagrangian
density).

We assume that the vector fields $A_{\mu }^{i}$ each take the form
\begin{equation}
A_{\mu }^{i}(x)=a_{\mu }^{i}(x)+n_{\mu }^{i}  \label{ab}
\end{equation}
when SBLS occurs; here the constant Lorentz four-vectors $n_{\mu }^{i}$ 
($i=1,...N$) are the VEVs of the vector fields. Substitution of 
the form (\ref{ab}) into the Lagrangian density (\ref{LN})
shows that the kinetic term for the vector fields $A_{\mu }^{i}$ 
translates into a kinetic term for the vector fields 
$a_{\mu }^{i}$ ($F_{\mu \nu }^{(A)}=F_{\mu \nu }^{(a)}$), while their mass
and interaction terms are correspondingly changed. Now we consider at 
first just infinitesimally small $n_{\mu }^{i}$ four vectors.
Furthermore we introduce a
stronger form of the non-observability of SBLS principle, requiring 
exact cancellations between non-Lorentz invariant terms of the same 
structure in the Lagrangian density $L(a_{\mu}^{i}+n_{\mu }^{i},\psi )$ 
for {\em any} set of infinitesimal vectors $n_{\mu }^{i}$. 
Then we define a new set of vector fields ${\bf a}_{\mu }^{i}$ 
by the infinitesimal transformation

\begin{equation}
a_{\mu }^{i}={\bf a}_{\mu }^{i}-\alpha ^{ijk}\omega ^{j}(x){\bf a}_{\mu}^{k}
\textrm{ , \quad }\omega ^{i}(x)=n_{\mu }^{i}\cdot x_{\mu }  \label{rot}
\end{equation}
which includes the above coupling constants $\alpha ^{ijk}$ and the linear
``gauge'' functions $\omega ^{i}(x)$. We require that the Lorentz
symmetry-breaking terms in the cubic and quartic self-interactions of the
vector fields ${\bf a}_{\mu }^{i}$, including those arising from their
kinetic terms, should cancel for {\em any} infinitesimal vector 
$n_{\mu}^{i}$. This condition is satisfied 
if and only if the coupling constants $\alpha^{ijk}${\bf \ }
and $\beta ^{ijkl}$ satisfy the following 
conditions ({\bf a}) and ({\bf b}):

({\bf a}) $\alpha ^{ijk}$ is totally antisymmetric (in the indices $i$, $j$
and $k)$ and obeys the structure relations:

\begin{equation}
\alpha ^{ijk}{\bf \equiv }\alpha ^{[ijk]}\equiv \alpha _{[jk]}^{i}
\textrm{ ,\quad }\left[ \alpha ^{i},\alpha ^{j} \right] = 
-\alpha ^{ijk}\alpha ^{k}  
\label{alg}
\end{equation}
where the $\alpha ^{i}$ are defined as matrices with elements 
$(\alpha^{i})^{jk}=\alpha ^{ijk}$.

({\bf b}) $\beta ^{ijkl}$ takes the factorised form:

\begin{equation}
\beta ^{ijkl}=-\frac{1}{4}\alpha ^{ijm}\cdot \alpha ^{klm}.  \label{fac}
\end{equation}
It follows from ({\bf a}) that the matrices $\alpha ^{k}$ form the adjoint
representation of a Lie algebra, under which the vector fields transform
infinitesimally as given in Eq.~(\ref{rot}). In the case when the matrices 
$\alpha ^{i}$ can be decomposed into a block diagonal form, there appears a
product of symmetry groups rather than a single simple group.

Let us turn now to the mass term for the vector fields in the Lagrangian 
$L(a_{\mu }^{i}+n_{\mu }^{i},\psi )$. When expressed in terms of the
transformed vector fields ${\bf a}_{\mu }^{i}$ (\ref{rot}) it contains SBLS
remnants, which should vanish, of the type:

\begin{equation}
(M^{2})_{ij}(\alpha ^{ikl}\omega ^{k}{\bf a}_{\mu }^{l}{\bf a}_{\mu }^{j}
\textrm{ }{\bf +a}_{\mu }^{i}n_{\mu }^{j}){\bf =}0  \label{mmm}
\end{equation}
Here we have used the symmetry feature $(M^{2})_{ij}=(M^{2})_{ji}$ for a
real Hermitian matrix $M^{2}$ and have retained only the first-order terms
in $n_{\mu}^{i} $. These two types of remnant have different structures and
hence must vanish independently. One can readily see that, in view of the
antisymmetry of the structure constants, the first term in Eq.~(\ref{mmm})
may be written in the following form containing the commutator of the
matrices $M^{2}$ and $\alpha ^{k}$:

\begin{equation}
\left[ M^{2},\alpha ^{k}\right]_{jl}\omega ^{k}{\bf a}_{\mu }^{l}
{\bf a}_{\mu }^{j}=0  \label{com}
\end{equation}
It follows that the mass matrix $M^{2}$ should commute with all the matrices
$\alpha ^{k}$, in order to satisfy Eq.~(\ref{com}) for all sets of ``gauge''
functions $\omega ^{i}=n_{\mu }^{i}\cdot x_{\mu }$. Since the matrices 
$\alpha ^{k}$ have been shown to form an irreducible representation of a
(simple) Lie algebra, Schur's lemma implies that the matrix $M^{2}$ is a
multiple of the identity matrix, ($M^{2})_{ij}={\bf M}^{2}\delta_{ij}$, thus
giving the same mass for all the vector fields. It then follows that the
vanishing of the second term in Eq.~(\ref{mmm}) leads to the simple
condition:
\begin{equation}
{\bf M}^{2}(n^{i}\cdot {\bf a}^{i})=0  \label{m}
\end{equation}
for any infinitesimal $n_{\mu }^{i}$. Since the Lorentz gauge condition 
($\partial _{\mu }{\bf a}_{\mu}^{i}=0$) has
already been imposed, we cannot impose extra gauge conditions of the 
type $n^{i}\cdot{\bf a}^{i}=n_{\mu }^{i}\cdot {\bf a}_{\mu }^{i}=0$. 
Thus, we are necessarily led to:

({\bf c}) massless vector fields, 
$(M^{2})_{ij}={\bf M}^{2}\delta_{ij\textrm{ }}=0$.

Finally we consider the interaction term between the vector and fermion
fields in the ''shifted'' Lagrangian density $L(a_{\mu }^{i}+n_{\mu
}^{i},\psi )$. In terms of the transformed vector fields 
${\bf a}_{\mu}^{i}$
(\ref{rot}), it takes the form

\begin{equation}
({\bf a}_{\mu }^{i}-\alpha ^{ijk}\omega ^{j}{\bf a}_{\mu }^{k}\textrm{ }
+n_{\mu }^{i}{\bf )\cdot }\overline{\psi }\gamma _{\mu }T^{i}\psi
\label{fer}
\end{equation}
It is readily confirmed that the Lorentz symmetry-breaking terms (the second
and third ones) can be eliminated, when one introduces a new set of fermion
fields ${\bf \Psi }$ using a unitary transformation of the type:

\begin{equation}
\psi =\exp \left[ iT^{i}\omega ^{i}(x)\right] 
{\bf \Psi }\textrm{ , \qquad }\omega^{i}(x)=n^{i}\cdot x  \label{ff}
\end{equation}
One of the compensating terms appears from the fermion kinetic term and the
compensation occurs for any set of ``gauge'' functions $\omega^i(x)$ if and
only if:

({\bf d}) the matrices $T^{i}$ form a representation of the Lie algebra with
structure constants $\alpha ^{ijk}$:

\begin{equation}
\left[ T^{i},T^{j}\right] = i\alpha ^{ijk}T^{k}.  \label{TTT}
\end{equation}
In general this will be a reducible representation but, for simplicity, we
shall take it to be irreducible here. This means that the matter fermions 
${\bf \Psi }$ are all assigned to an irreducible multiplet determined by 
the matrices $T^{i}$. At the same time, the unitary transformation (\ref{ff})
changes the mass term for the fermions to

\begin{equation}
\overline{{\bf \Psi }}\left( m+i\omega ^{k}\left[ m,T^{k}\right] 
\right){\bf \Psi }  
\label{k}
\end{equation}
The vanishing of the Lorentz non-invariant term (the second one) in 
Eq.~(\ref
{k}) for any set of ``gauge'' functions $\omega^i(x)$ requires that the
matrix $m$ should commute with all the matrices $T^{k}$. According to
Schur's lemma, this means that the matrix $m$ is proportional to the
identity, thus giving:

({\bf e}) the same mass for all the fermion fields within the
irreducible multiplet determined by the matrices $T^{i}$:
\begin{equation}
m_{rs}={\bf m}\delta _{rs}.
\end{equation}
In the case when the fermions are decomposed into several irreducible
multiplets, their masses are equal within each multiplet.

Now, collecting together the conditions ({\bf a})-({\bf e}) derived from the
non-observability of the SBLS for {\em any} set of infinitesimal vectors 
$n_{\mu }^{i}$ applied to the general Lagrangian density (\ref{LN}), we
arrive at a truly gauge invariant Yang-Mills theory for the new fields 
${\bf a}_{\mu }^{i}$ and ${\bf \Psi }$:

\begin{equation}
L_{YM}={\bf -}\frac{1}{4}{\bf F}_{\mu \nu }^{i}{\bf F}_{\mu \nu }^{i}+
i\overline{{\bf \Psi }}\gamma \partial {\bf \Psi -m}\overline{{\bf \Psi }}
{\bf \Psi +ga}_{\mu }^{i}\overline{{\bf \Psi }}{\bf \gamma }_{\mu }
{\bf T}^{i}{\bf \Psi }  \label{fin}
\end{equation}
Here ${\bf F}_{\mu \nu }^{i}=\partial_\mu {\bf a}_{\nu }^{i}  -
\partial_{\nu }{\bf a}_{\mu }^{i} {\bf +ga }^{ijk}{\bf a}_{\mu }^{j}
{\bf a}_{\nu}^{k}$ and ${\bf g}$ is a universal gauge coupling 
constant extracted from the corresponding matrices 
$\alpha ^{ijk}={\bf ga }^{ijk}$ and $T^{i}={\bf gT}^{i}$.

Let us now consider the generalisation of the vector field VEVs from
infinitesimal to finite background classical fields $n_{\mu }^{i}$.
Unfortunately one cannot directly generalise the SBLS form (\ref{ab}) to all
finite $n_{\mu }^{i}$ vectors. Otherwise, the $n_{\mu }^{i}$ for the
different vector fields might not commute under the Yang-Mills symmetry and
might point in different directions in Lorentz space, giving rise to a
non-vanishing field strength $F_{\mu \nu }^{k}$ in the corresponding vacuum.
Such a vacuum would not be Lorentz invariant, implying a real physical
breakdown of Lorentz symmetry. This problem can be automatically avoided if
the finite SBLS shift vector $n_{\mu }^{i}$ in the basic equation (\ref{ab})
takes the factorised form $n_{\mu }^{i}=n_{\mu }\cdot f^{i}$ 
where $n_{\mu}$
is a constant Lorentz vector as in the Abelian case, while $f^{i}$ ($%
i=1,2,...N$) is a vector in the internal charge space. Using the Lagrangian
density (\ref{fin}) derived for infinitesimal VEVs, it is now
straightforward to show 
that there will be no observable effects of SBLS for
{\em any} set of finite factorised VEVs $n_{\mu }^{i}=n_{\mu }\cdot f^{i}$.
For this purpose, we generalise
Eq.~(\ref{rot}) to the finite transformation:

\begin{equation}
a_{\mu }\cdot \alpha =\exp [(\omega \cdot \alpha )]{\bf a}_{\mu }\cdot
\alpha \exp [-(\omega \cdot \alpha )]  \label{finiterot}
\end{equation}

In conclusion, we have shown that gauge invariant Abelian and non-Abelian
theories can be obtained from the requirement of the physical
non-observability of the SBLS rather than by using the Yang-Mills gauge
principle. Thus the vector fields become a source of the symmetries, rather
than local symmetries being a source of the vector fields as in the usual
formulation. Imposing the condition that the Lorentz symmetry breaking be
unobservable of course restricts the values of the coupling constants and
mass parameters in the Lagrangian density. These restrictions may naturally
also depend on the direction and strength of the Lorentz symmetry breaking
vector field VEVs, whose effects are to be hidden. This allows us a choice
as to how strong an assumption we make about the non-observability
requirement. Actually, in the Abelian case, we just assumed this
non-observability for the physical vacuum (\ref{f}) that really appears.
However we needed a stronger assumption in the non-Abelian case: the SBLS is
unobservable in any vacuum for which the vector fields have VEVs of the
factorised form $n_{\mu }^{i}=n_{\mu }\cdot f^{i}$. This factorised form is
a special case in which the $n_{\mu }^{i}$ commute with each other.

We did not specify here mechanisms which could induce the 
SBLS---rather we studied general consequences 
for the possible dynamics of the
matter and vector fields, requiring it to be physically unobservable. We 
address this and other related questions elsewhere\cite{long}.

We are indebted to Z. Berezhiani, H. Durr, O. Kancheli, K. Maki, 
P. Minkowski and D. Sutherland for stimulating discussions and useful remarks.

\end{document}